\documentclass[useAMS,usenatbib]{mn2e}
\usepackage{graphicx}                                            
\usepackage{times}
\usepackage{float}
\usepackage{rotating}
\usepackage{epstopdf}

\def\lesssim{\mathrel{\hbox{\rlap{\hbox{\lower4pt\hbox{$\sim$}}}\hbox{$<$}}}}
\let\la=\lesssim
\def\gtrsim{\mathrel{\hbox{\rlap{\hbox{\lower4pt\hbox{$\sim$}}}\hbox{$>$}}}}
\let\ga=\gtrsim

\def\lx{$L_\mathrm{X}$}

\def\ergs{erg s$^{-1}$}
\title[Inclination effects in black hole transients ]{Inclination and relativistic effects in the outburst evolution of\\ 
black hole transients}
\author[Mu\~noz-Darias et al.]{T.~Mu\~noz-Darias$^{1}$, M.~Coriat$^{1,2}$, D.~S.~Plant$^{1}$, G.~Ponti$^{3,1}$, R.~P.~Fender$^{1}$ and R.~J.~H.~Dunn$^{4}$ \\
$^{1}$School of Physics and Astronomy, University of Southampton, Southampton, Hampshire, SO17 1BJ, United Kingdom\\
$^{2}$Department of Astronomy, University of Cape Town, Private Bag X3, Rondebosch, 7701, South Africa\\
$^{3}$Max Planck Institute fur Extraterrestriche Physik, 85748, Garching, Germany\\
$^{4}$ Formally at the Excellence Cluster 'Universe', Technische Universität Munchen, Boltzmannstrasse 2, D-85748 Garching, Germany}

\begin{document}
\maketitle
\begin{abstract}
We have systematically studied the effect of the orbital inclination in the outburst evolution of black hole transients. We have included all the systems observed by the Rossi X-ray timing explorer in which the thermal, accretion disc component becomes strongly dominant at some point of the outburst. Inclination is found to modify the shape of the tracks that these systems display in the colour/luminosity diagrams traditionally used for their study. Black hole transients seen at low inclination reach softer spectra and their accretion discs look cooler than those observed closer to edge-on. This difference can be naturally explained by considering inclination dependent relativistic effects on accretion discs.\\      
\end{abstract}
\section{Introduction}
Stellar-mass black holes are mostly found as black hole transients (BHTs), a flavour of X-ray binaries in which a $\sim$ 5--15 M$_{\odot}$ black hole is accreting from a low mass donor star. These systems spend most of their lives in quiescence, showing occasional outbursts where their X-ray emission increases by up to eight orders of magnitude, reaching X-ray luminosities in the range \lx$\sim10^{36}$--$10^{39}$ \ergs. The evolution of BHTs during these activity epochs is usually studied using the hardness-intensity diagram (HID; \citealt{Homan2001}), which traces the behaviour of the X-ray luminosity and energy spectrum. The presence of various X-ray states (e.g. ~\citealt{Nowak1995}, \citealt{Remillard2006b}, \citealt{vanderklis2006},  \citealt{Belloni2011}, \citealt{Fender2012}) possibly connected to different accretion regimes, becomes evident in this diagram, where a standard pattern is recurrently repeated by the vast majority of these objects.\\  
A fast rise at $\sim$ constant X-ray colour\footnote{defined as the flux ratio between a hard and a soft energy band.} is observed at the beginning of the outburst. The X-ray spectrum is dominated by high energy emission that can be modelled by a cutoff (around $\sim 100$ keV) power-law and it is thought to be produced in the innermost part of the accretion flow. A cooler ($\lesssim$ 2 keV) thermal component is sometimes observed depending on the system and the low-energy instrumental cutoff. At some point, the BHT leaves this so-called $hard$ state and moves towards softer energy spectra as the thermal component becomes more dominant and the power-law emission weaker. This transition occurs at approximately constant luminosity in contrast to the major changes observed during the hard state. The thermal component associated with a geometrically thin, optically thick accretion disc (\citealt{Shakura1973}) dominates the energy spectrum during the $soft$ state, where BHTs typically spend most of the outburst. The luminosity gradually decreases during this state until reaching values around $\sim 2\%$ of the Eddington limit (\citealt{Maccarone2003}), at which point,  a soft-to-hard transition takes place. Subsequently, the system starts a decay towards quiescence following a similar, vertical HID track to that observed during the initial rise.\\
The above q-shaped pattern has been seen in the HID of a number of BHT, especially after the launch of the \textit{Rossi X-ray timing explorer} (\textit{RXTE}) in 1995 (see e.g.  \citealt{Fender2009}) . However, whereas the hard state is virtually always observed, some systems never leave it or display occasional outbursts in which a proper soft state is never reached (e.g. \citealt{Motta2010}, \citealt{Coriat2011} for H1743-322; see also \citealt{Brocksopp2004}). Even if one only considers the systems that follow the \textit{canonical} HID pattern, some differences between them can still be noticed. Recently, \cite{Ponti2012} studied the presence/absence of accretion disc winds in some BHT seen at high and low inclination, respectively. A quick inspection of the HIDs presented in that work suggest that the inclination could have an effect on the shape of the tracks. Here, we investigate this problem in a general way using data from all the BHT observed during the \textit{RXTE} era.

\begin{table*}
\begin{center}

\caption{List of black hole transients and outbursts included in this work}
\label{log}
\begin{tabular}{c c c c c}
\hline
\hline
System & Outbursts & Inclination (deg) & Distance (kpc) & Orbital period (h) \\
\hline
4U 1543-47 & 2002 & $20.7 \pm 1.5$ (1) & $7.5 \pm 0.5$ (1) & 27 (1)\\
XTE J1650-500 & 2001 & $\gtrsim$ 47 (2) & $2.6\pm 0.7$ (2) & 7.63 (3)\\
GX 339-4 & 2002, 2004, 2007 &  & $\geq$ 6 (4) & 42.2 (5) \\
XTE J1720-318 & 2003 &  &  & \\
XTE J1752-223 & 2009 & $\lesssim 49$ (6; a) & $6\pm2$ (7) & $\lesssim 6.8$ (7) \\
XTE J1817-330 & 2006 & 	& $5\pm4$ (8) & \\
XTE J1859+226 & 1999 & $\gtrsim 60$ (9; b) & $6.2 \pm 1.8$ (10); $\sim 14$ (9) & 6.68 (9)  \\
\hline
XTE J1550-564 & 1998 & $74.7 \pm 3.8$ (11) & $4.38^{+0.58}_{-0.41}$ (11) & 37.01 (11)\\
4U 1630-47 & 2004, 2006, 2007  & Dipping (12,13)& $\sim 10$ (14) & $\gtrsim 13$ (14)\\
GRO J1655-40 & 1996, 2005 & $70.2\pm 1$ (15); Dipping (13) & $3.2\pm0.2$ (16) & $62.4\pm0.6$ (17)\\ 
H1743-322 & 2003, 2004 &  $75\pm3$ (18,19; a); Dipping (20)& $8.5\pm0.8$ (19)\\
\hline
\hline
\end{tabular}
\end{center}

\begin{flushleft}
(a) Assuming that the radio jet is perpendicular to the orbital plane.\\
(b) Inclination $\sim$ 60 degrees if accretion disc does not contribute to the optical luminosity in quiescence.\\
\textsc{REFERENCES:} 
(1)\citet{Orosz2003};~(2)\citet{Orosz2004};~(3) \citet{Homan2006};~(4)\citet{Hynes2004}; (5)\citet{Hynes2003};~(6)\citet{Miller-Jones2011};~(7)\citet{Ratti2012};~(8)\citet{Sala2007}; ~(9)\citet{Corral-Santana2011};~(10)\citet{Hynes2002};~(11)\citet{Orosz2011};~(12)\citet{Tomsick1998}; ~(13)\citet{Kuulkers1998}; ~(14)\citet{Augusteijn2001}; ~(15)\citet{Greene2001}; ~(16)\citet{Hjellming1995};~(17) \citet{Bailyn1995}; ~(18)\citet{Corbel2005}; ~(19)\citet{Steiner2012}; ~(20)\citet{Homan2005b} 

\end{flushleft}

\end{table*}

\section{Methodology and results}
The goal of this work is to perform a systematic study on the effect that the orbital inclination has in the outburst evolution of BHTs. Our approach to this problem relies on two main assumptions: (i) there is no intrinsic difference between low inclination and high inclination systems and (ii) the intrinsic outburst evolution of our sample of objects is similar, the line-of-sight towards the Earth being the main difference between them. The first assumption cannot be proved, but we think it is reasonable, and no systematic difference is obvious in the main properties of our sample (e.g. orbital parameters; table~\ref{log}). In an attempt to accomplish the second statement we have selected only those sources and outbursts which have been densely monitored and have shown strongly disc dominated soft states. These outburst tend to be the brightest and longest for those systems showing both \textit{standard} and \textit{only-hard} outburst (see e.g. Dunn et al. 2010; hereafter D10). However, there are sources showing solely long, relatively bright, hard outburst that are not included in our sample (e.g. Swift~J1753.5-0127; \citealt{Soleri2013}).

\subsection{The sample}
For this study we make use of the spectral fits presented by D10, where all the BHT observed by \textit{RXTE} until 2009 are analysed. These fits are computed within the spectral range 3--250 keV using Power-law, broken power-law or power law+ disc (see D10 for further details). For those with discs, a Newtonian multi-blackbody spectral model (\textsc{diskbb}; \citealt{Mitsuda1984}) is used. \par
From the above database we have used the power-law fraction luminosity diagrams presented in D10 to evaluate the disc contribution to the $total$ flux (power-law + disc) for each object and activity period (see also section \ref{s:plf}). For this work, we have arbitrarily chosen only those outbursts in which the thermal component accounts for at least 95\% of the total flux at some point of the outburst. Since power-law fraction measurements can have errors of a few percent, this constraint ensures that we are only considering outbursts with stable, disc dominated soft states. In addition, We have included 204 observations corresponding to XTE~J1752-223; a BHT discovered after D10 (\citealt{Markwardt2009}, \citealt{Munoz-Darias2010}). These observations were fitted using the same methodology reported in D10, giving results consistent with previous studies (\citealt{Shaposhnikov2010a}). The list of objects and activity periods used in these work is presented in table \ref{log}. All the BHTs observed outside the hard states and extensively monitored by \textit{RXTE} are included with the exception a few sources. XTE~J1748-288 and MAXI J1659-152 do not show strongly disc dominated soft states (power law contribution is always higher than 5\%; see also D10, \citealt{Munoz-Darias2011b}). MAXI~J1543-564 is omitted because due to its low count rate (see \citealt{Stiele2012}) the spectral analysis gets very complex and, in any case, our fits are consistent with power-law contributions higher than 5\%. XTE~J1652-453 is excluded since \textit{RXTE} observations started late in the soft state. However, we note that for the latest a low inclination has been proposed through X-ray reflection fitting and its HID mimics that of GX~339-4 (\citealt{Hiemstra2011}). Finally, GRS 1915+105 is also excluded given its unusual outburst evolution behaviour.\par
The eleven systems and seventeen outbursts are divided into two groups. Four (displaying eight outbursts) are considered high inclination sources ($i\gtrsim 70$  degrees; bottom group in table \ref{log}). For three of these four sources this classification is inferred by the presence of absorption dips in X-rays together with the detection of a high inclined radio jet (H 1743-322) and optical measurements in quiescence (GRO J1655-40). X-ray dips have not been detected in XTE J1550-564 but optical observations proved the edge-on line-of-sight for this source. The remaining seven systems (accounting for nine outbursts) are classified as low inclination objects (top group in table \ref{log}). Not all the systems belonging to this second category have an accurate inclination measurement, but the non detection of high inclination features after years of intense monitoring clearly points towards viewing angles $<70^{\circ}$ [e.g. GX 339-4; note that X-ray reflection fitting techniques also favour low inclination values (e.g. \citealt{Reis2008}) for this system]. However, we note that at least one system (XTE J1859+226) could have a moderately high inclination ($\gtrsim60^{\circ}$).

\begin{figure*}
\centering
\includegraphics[width= 19cm,height=22cm]{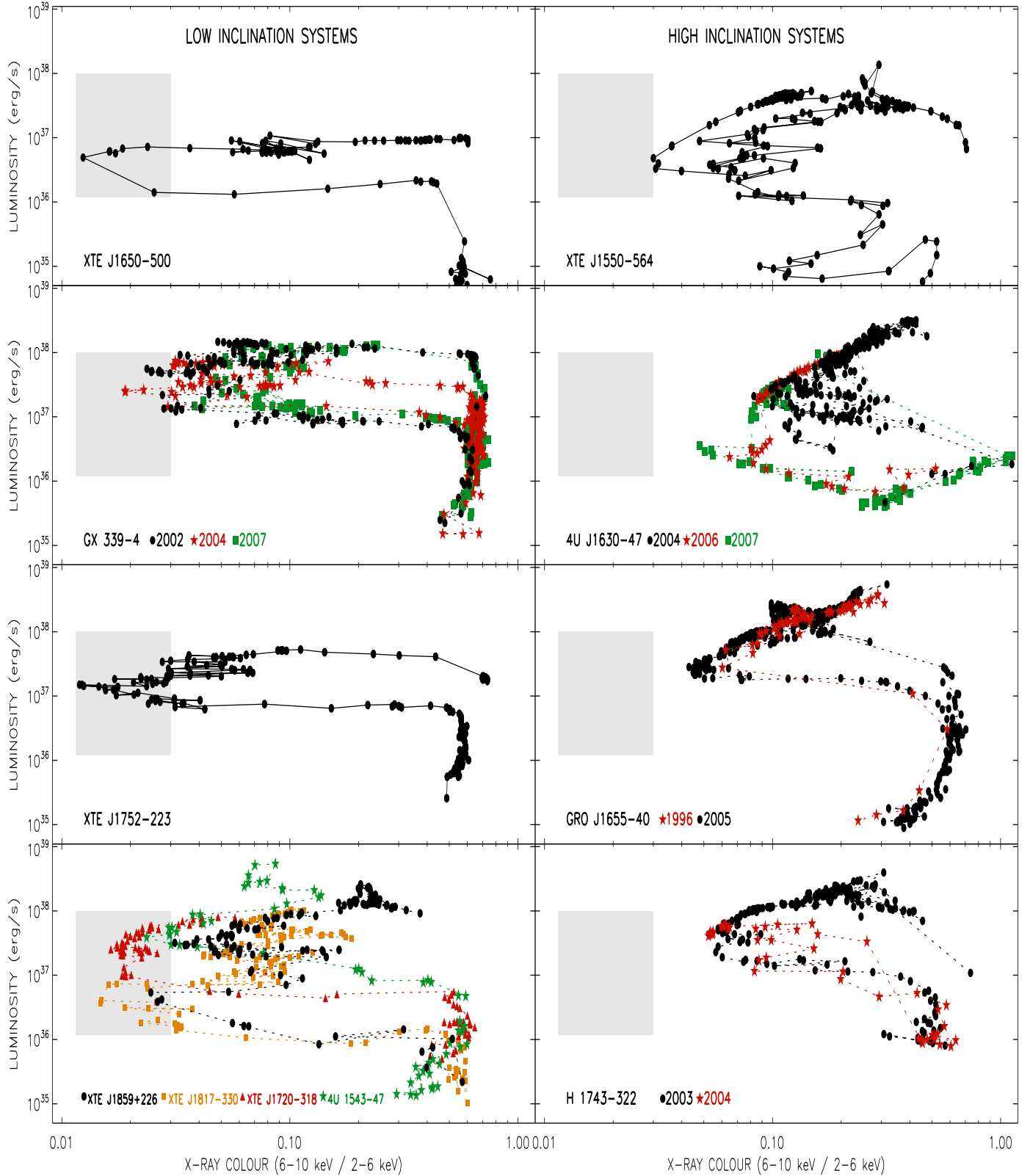}

\caption{Hardness intensity diagrams for low (left panels) and high inclination sources (right panels). Luminosity is computed using unabsorbed fluxes in the range 2--16 keV and the distance measurements reported in table \ref{log} (8 kpc if unknown). Grey rectangles mark the X-ray colour-luminosity region reached solely by low inclination sources. Solid line (dashed in case of more than one outburst) join contiguous observations. }
\label{hids}
\end{figure*}

\begin{figure}
\centering
\includegraphics[width= 8.5cm,height=8cm]{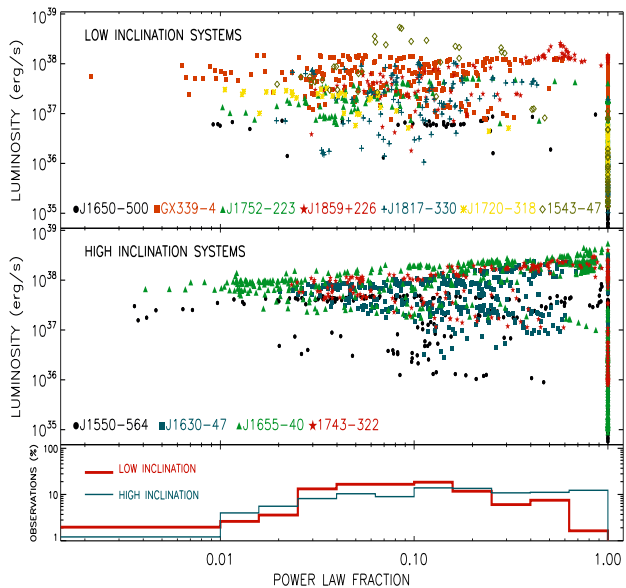}
\caption{Power-law fraction luminosity diagrams for low (top panel) and high inclination sources (middle panel). PLF is computed following eq.\ref{eplf} and luminosity is the same as in Fig.\ref{hids}. Bottom panel: fractional distribution of observations along ten (logarithmically) uniform bins within the range  $0.01<PLF<1$ (i.e. only observations showing a disc component are considered). All the observations with $PLF<0.01$ are showed in one single bin.}
\label{plf}
\end{figure}

\subsection{Hardness intensity diagrams}
We computed HIDs for each outburst of each BHT. Luminosities are calculated within the range 2-16 keV using the distance measurements reported in table \ref{log} and assuming that the sources emit isotropically. If the distance is unknown we use 8 kpc. X-ray colours are computed by dividing the absorption corrected flux in the band 6-10 keV by that in the range 2-6 keV (i.e. the higher the X-ray colour, the harder the spectrum). The resulting HIDs for low (left panels) and high (right panels) inclination systems are presented in Fig. \ref{hids}. Some differences seem to be present between the two groups:
\begin{itemize}
\item Low inclination systems are systematically softer during the soft state. Indeed, no edge-on system become softer than X-ray colour $\sim 0.03$ and therefore than any face-on BHT (see grey area in Fig. \ref{hids}). 
\item In low inclination systems luminosity remains approximately constant along the hard-to-soft transition, and decreases roughly \textit{vertically} (i.e. at constant X-ray colour) during the soft state.  This results in HID tracks with \textit{square} shapes. At high inclination we see that the luminosity decreases monotonically all the way from the middle of the hard-to-soft transition to the softest points. This yields HIDs with more \textit{triangular} tracks.
\end{itemize}

\subsection{Power-law fraction and Disc temperature}\label{s:plf}
As a second step we computed power-law fraction luminosity diagrams (\citealt{kording2008}) for both groups of objects (see top and middle panels in Fig. \ref{plf}). Luminosity is defined as in Fig. \ref{hids} and power-law fraction ($PLF$) is obtained following D10:
\begin{equation}
\label{eplf}
PLF =\frac{L_{PL}}{L_{PL}+L_{D}}
\end{equation}
where $L_{PL}$ and $L_D$ are the power-law and disc luminosities in the ranges 1--100 keV and 0.001--100 keV, respectively. Therefore, $PLF=1$ corresponds to observations in which the disc, thermal component is not detected. We note that using $L_D$ from 0.001 keV instead of e.g. 1 keV is useful to increase the dynamical range of PLF and does not affect either our selection of sources or outbursts. For each group of systems, a histogram showing the distribution of observations across several $PLF$ bins is presented in the bottom panel of Fig. \ref{plf}. Here, we have considered only those observations where the two spectral components are present ($PLF<1$). They account for 48\% and 70\% of the total for low and high inclination sources, respectively. Some conclusions can be derived from Fig \ref{plf}:

\begin{figure}
\centering
\includegraphics[width= 8.5cm,height=7cm]{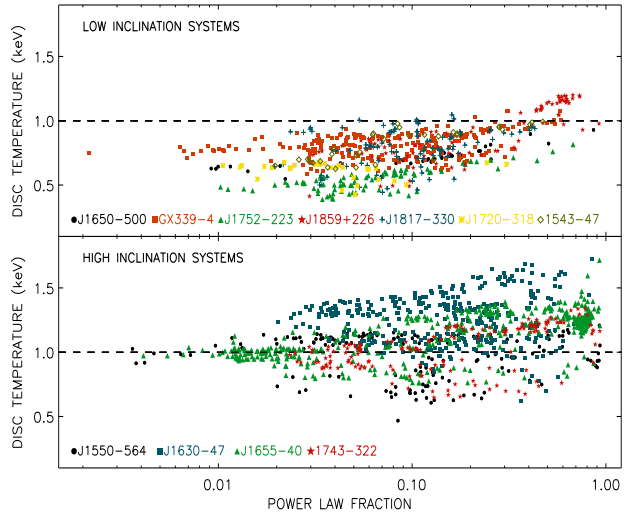}
\caption{Disc temperature as a function of the PLF for low (top panel) and high inclination sources (bottom panel). PLF is computed in the same way as in Fig.\ref{plf}}
\label{tdisc}
\end{figure}  

\begin{itemize}
\item The shapes described by the high inclination sources look more triangular, as seen in the HIDs as well.
\item All the systems reach at some point power-law fractions well below the 0.05 level imposed to the sample, which argues that the sample of objects included in our study is not biassed by any viewing angle effect. No systematic difference between low and high inclination sources seems to be present when comparing the smallest PLF values observed. In other words, there is no evidence supporting that one population of objects shows soft states more disc dominated than the other. 
\item The distribution of observations for $PLF<0.6$ is similar for both groups of objects. 
\item The number of observations in the range $0.6<PLF<1$ is extremely small for low inclinations systems, whereas is above 10\% of the total for BHT closer to edge-on. We think that this difference is significant although we note that three of the seven low inclination systems (accounting for three outbursts) were not monitored during the early stage of the hard-to-soft transition. An exception to this behaviour seems to be again XTE J1859+226 (stars in top panel), which look more similar to edge-on systems. Indeed, the vast majority of the low inclination observations above $PLF\sim0.6$ correspond to this source.
\end{itemize} 

In Fig. \ref{tdisc} we present the evolution of the disc temperature ($T_{obs}$) as a function of the $PLF$. $T_{obs}$ is the temperature at the inner disk radius directly obtained from the Newtonian \textsc{diskbb} model (however, see section \ref{tobs-teff} ). We find that low inclination sources systematically have cooler discs with $T_{obs} \lesssim 1$ keV (dotted line in Fig. \ref{tdisc}). If for each BHT we take the 5 highest $T_{obs}$ within the range $0.01<PLF<0.1$, we find low inclination systems moving in the range 0.60--1.00 keV  around $0.80 \pm 0.13$ keV. For edge-on sources, the same test gives $T_{obs}$=$1.22~\pm~0.14$ keV and temperatures within the interval 1.06--1.42 keV. This $PLF$ range is convenient because (i) it is densely covered by all the objects and (ii) the disc component is dominant and easier to disentangle from the power-law emission. Using this method we estimate that high inclination BHTs look a factor $1.53 \pm 0.31$ hotter than low inclination sources. We note that the above method ensures that all the systems are equally weighted in the average. On the other hand, significant variations of either the number of temperature values used to calculate the average or the PLF range considered produce changes well within the errors of latest factor. Finally, we also looked at the distribution of power-law indexes as a function of the PLF. Power-law index values roughly span the same range for both groups, and no obvious difference is present in the data. We find similar results when looking at the power-law normalization values, although we note that the large errors associated to the distance estimates do not allow us to fully explore this issue. 

\section{Discussion}
\label{discussion}  
We have shown that the system inclination has a clear impact on the observed outburst evolution of BHT. We observe that during the nine outbursts displayed by the low inclination sources the X-ray colour becomes softer than during any of the eight outbursts associated with high inclination systems. The only exception to this behaviour is the 2007 outburst of GX 339-4, which is slightly harder than the 1998 outburst of XTE J1550-564 (see Fig. \ref{hids}). However, the high energy, power-law component still accounts for $\sim 5$\% of the flux during the softest observation of the former (i.e. exactly the limit to be part of this study), whereas the disc is the only spectral component present in the equivalent observation of XTE J1550-564 ($PLF \sim 0.003$; Fig. \ref{plf}). On the other hand, we systematically measure higher disc temperatures in high inclination sources (Fig. \ref{tdisc}). This difference varies from system to system, but edge-on sources look on average a factor $\sim 1.5$ hotter than those seen at lower inclination. Since the disc is nearly the only spectral component present during the softest epochs, hotter discs naturally explain the harder X-ray colours observed in the high inclination sources. They also explain why in these sources the disc is detected earlier in the outburst (at $PLF>0.6$; Fig. \ref{plf}) above the low energy cutoff of \textit{RXTE}. In agreement with this, 70 \% of the observations of high inclination systems show a thermal component, which is detected only in 48\% of the low inclination BHTs. However, we note that the latter number could be biased by sampling effects. For instance, the initial rise along the hard state was not detected (i.e. \textit{RXTE} observations started later) for the vast majority of sources and outburst, being solely observed for the three outbursts of GX~339-4 (low inclination), one of them (2004) being particularly long and extensively monitored (see Fig. \ref{hids}). Thus, the fraction of observations of low inclination systems where only the power-law component is detected could be easily affected by this.

\subsection{Observed vs. effective disc temperature}\label{tobs-teff}    
The spectral fits employed in this work make use of a standard Newtonian multi-blackbody spectral model (\textsc{diskbb}; \citealt{Mitsuda1984}). This simple model is particularly useful when performing systematic studies that involve many observations of many systems, as that presented in D10. However, it is known that the correspondence between the \textsc{diskbb} parameters and the actual physical parameters can be complex and indirect (e.g. \citealt{Merloni2000}). The observed disc temperature, $T_{obs}$, is in reality:
\begin{equation}
\label{tobs}
T_{obs}=f_{GR}(i,a_{*})\times  f_{col} \times \xi~T_{eff}
\end{equation}
where $T_{eff}$ is the effective temperature at the innermost disc radius and $f_{col}$ is known as the spectral hardening term (see below). \textsc{diskbb} assumes that temperature is always proportional to $ r^{- \frac{3}{4}}$ (r being the radius). This deviates at small radii from that obtained if applying a zero-stress inner boundary condition, which might affect the obtained $T_{obs}$ . A $\xi$ factor, that is found to be as small as $\sim 1.04$ by \cite{Gierlinski1999}, is introduced to compensate this deviation.\\ 
The first term in eq. \ref{tobs}, $f_{GR}(i,a_{*})$, accounts for relativistic effects in a strong gravitational potential (\citealt{Cunningham1975}, \citealt{zhang1997}). This correction assumes that the inner disc is at the innermost stable circular orbit (ISCO), and depends on the orbital inclination, $i$, and the dimensionless BH spin parameter, $a_{*}$ (where $0 \leq a_{*} \leq 1$). A non-rotating (Schwarzschild) black hole corresponds to $a_{*}=0$, whereas a maximally rotating (extreme Kerr) black hole implies $a_{*}=1$. Some aspects must be considered regarding the $f_{GR}$ term: 
\begin{itemize}
\item gravitational redshift is expected to affect the emission arising from the innermost regions of the accretion disc. Therefore, an observer at infinity will perceive the light shifted towards lower frequencies (i.e. longer wavelengths), making the disc to look cooler. Additionally, matter is moving at a few tenths of the speed of light in the innermost part of the accretion disc and relativistic Doppler effects are relevant. In particular, at low inclinations the transverse Doppler effect term become important, causing a further redshift. Therefore, $f_{GR}$ is less than unity for sources close to face-on.

\item As $i$ increases, other effects must be taken into account. In particular, one should take into account that at these relativistic velocities light becomes partially focussed towards the direction of movement (Doppler beaming). This enhances the light arising from the material moving towards the observer as compared to that emitted from the receding flow. A net blueshift effect opposite  to that resulting from the gravitational redshift is therefore expected at high inclination. This makes $f_{GR}$ as large as $1.35$--$1.66$ for $i=90^{\circ}$. This range in $f_{GR}$ is a consequence of its dependence with $a_{*}$. The larger the spin, the smaller the innermost disc radius and the more important relativistic corrections are.
 
\end{itemize}

We note that fully relativistic disc models that consider all the above effects have clearly shown the expected dependence of the observed disc spectrum with the inclination angle (see \citealt{Li2005} and section 3.2). On the other hand, these effects are balanced for orbital inclinations close to $\sim 75^{\circ}$ (\citealt{zhang1997}). Thus,  $f_{GR}$ becomes close to unity regardless of the spin parameter for that viewing angle. This prediction is consistent with our results. We find that the three systems with inclination measurements around that value, namely, XTE~J1550-564, XTE~J1655-40 and H1743-322 show very similar $T_{obs}$ at low PLF values (i.e. when the disc dominates; bottom panel in Fig. \ref{tdisc}). On the contrary, the range of inclinations present within the low inclination objects  is expected to be much broader just from the way we have divided our sample. Looking at the distribution of temperatures for these sources at low PLF values ($PLF \lesssim 0.05$; top panel in Fig. \ref{tdisc}) we can roughly split the systems in three groups:
\begin{enumerate}
\item XTE~J1752-223 shows the lowest $T_{obs}$ and it is also the softest source of our sample (see Fig. \ref{hids}). This agrees with the low inclination suggested by the orientation of its radio jet ($ i \lesssim 48^{\circ}$; \citealt{Miller-Jones2011})
\item Three systems have intermediate temperatures: 4U~1543-47, XTE J1650-500 and XTE J1720-318. The first two seem to have different inclinations (see table \ref{log}) and the inclination of the last one is unknown.
\item GX~339-4, XTE J1817-330 and XTE J1859+226 show the highest temperatures of their group. We only have an inclination constraint for XTE J1859+226, which is suggested to be among the highest (if not the highest) of the low inclination sources ($i\gtrsim60^{\circ}$; \citealt{Corral-Santana2011}). This is consistent with the system displaying the hardest soft state of its class.  
\end{enumerate}
We remark that in the above discussion we have focussed on the softest observations, where $T_{obs}$ measurements are more reliable (the disc is virtually the only spectral component), the range of accretion rates displayed by all the systems might be narrower than that for other stages of the outburst and the condition of inner disc being at ISCO expected to be satisfied. However, we note that relatively similar BH masses, spins and accretion rates are needed to compare $T_{obs}$ measurements purely based on the distribution of orbital inclinations.\\
On the other hand, if we consider that the low inclination sources of our sample have, on average, an inclination around $\sim 40^{\circ}$, we should expect $T_{obs}$ to be a factor 1.2--1.8 smaller (depending on the spin parameter) for low inclination than for edge-on BHTs \footnote{Values correspond to $a_{*}$ equal 0 and 0.998, respectively. We use corrections from \cite{zhang1997} for $i\sim 41.4^{\circ}$ (low inclination) and $75.5^{\circ}$ (high inclination).}. This is in remarkable agreement with the $\sim 1.5\pm 0.3$ factor we  estimate from our sample of BHTs. In the case that the scatter in this factor is dominated by inclination effects, our results do not favour all the BHs of our sample having either $a_{*}=0$ or $a_{*}=1$. If that were the case, one would expect factors distributed around 1.2 or 1.8, respectively. This suggest that at least some of the BHs of our sample are spinning, and seems consistent with the broad range of spin values reported in the literature (see \citealt{McClintock2013} and \citealt{Reynolds2013} for recent reviews) . However, we caution the reader that the distribution of inclinations for the low inclination systems is not constrained and that systematic differences in e.g. compact object mass or accretion rate could bias this result. Additionally, we note that in the above discussion a good alignment between the BH spin and the binary orbit is assumed. \par 

Light bending effects also modify the amount of disc flux reaching a distant observer. For spin values close to unity this compensates the larger projected area that low inclination systems ($i \sim$ 40) have as compared to edge-on sources ($i \sim 75$), and even in the case of $a_{*}=0$ the expected difference is reduced from a factor $\sim 3$ (Newtonian case) to a factor $\sim 2$ (\citealt{zhang1997}). Our database and fits are probably not sensitive enough to test those variations. Moreover, given the uncertainties in the distances, it is difficult to study systematics in the disc luminosities within a factor of 2. However, this issue could also play a role when comparing HID tracks for low and high inclination sources (see below).\par

The second correction term present in eq. \ref{tobs} is known as the spectral hardening factor ($f_{col}$; see \citealt{Shimura1995}). Given the high temperatures reached in the inner part of the accretion disc, its emission is not entirely dominated by absorption opacity (as expected for black body emission) and scattering processes become important. Since the same amount of energy must be still dissipated, the observed temperature is seen to increase by a factor $f_{col}$ (see also \citealt{Merloni2000}; \citealt{Gierlinski2004}). \cite{Shimura1995} found $f_{col}$ to vary between $\sim$ 1.6~--~2 for wide ranges of BH masses and accretion rates. However, more detailed calculations including non-LTE effects performed by \cite{Davis2005} report smaller values in the range $\sim$ 1.4~--~1.6. A possible, small ($<10$\%) dependence of $f_{col}$ with the inclination due to limb darkening  effects is also suggested. In any case, assuming that there is no systematic difference in BH masses and accretion rates for high and low inclination objects, $f_{col}$ is not expected to produce the variation in $T_{obs}$ that we see. 

\begin{figure}
\centering
\includegraphics[width= 8.5cm,height=7cm]{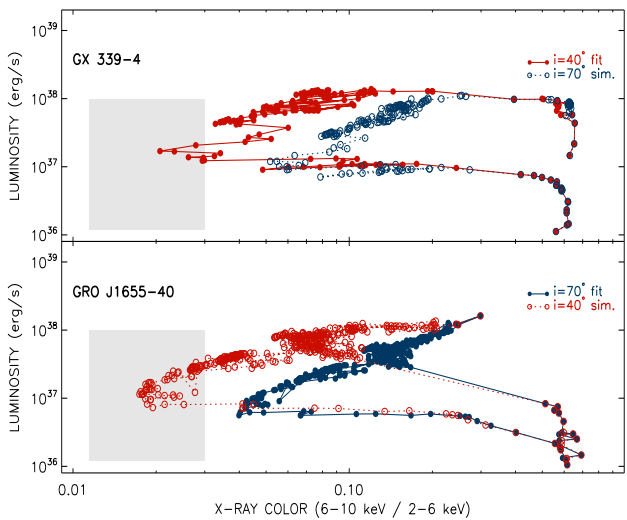}
\caption{Hardness intensity diagrams for GX~339-4 (2002 outburst; top panel) and GRO~J1655-40 (2005 outburst; bottom panel) derived using a fully relativistic \textsc{kerrbb} model for the accretion disc (see text). Solid lines joining filled circles represent the actual fit to the data whereas open circles correspond to the predicted values in case the inclination is switched from $40^{\circ}$ to $70^{\circ}$ and vice versa. A BH mass of 7 $M_{\odot}$ and a spin value of $a_*=0.5$  has been used in both cases. The grey region is the same as in Fig. \ref{hids}.}
\label{kerrbb}
\end{figure} 

\subsection{Fully relativistic disc models}
The relativistic effects discussed above naturally reproduce the observed difference in disc temperature. Discs which appear hotter emit harder X-ray emission, which gives an explanation for the harder soft states present in the edge-on sources. In order to quantify more precisely this effect, and see how it might affect other stages of the outburst, we have made use of a fully relativistic accretion disc model. A model considering the multi-temperature blackbody spectrum of a thin accretion disk around a Kerr black hole (\textsc{kerrbb}; \citealt{Li2005}) was applied to one source of each group, namely GX~339-4 (low inclination; 2002) and GRO~J1655-500 (high inclination; 2005). The test was carried out in two steps as follows: 
\begin{enumerate}
\item we fitted all the observations corresponding to each outburst replacing the Newtonian model (\textsc{diskbb}) by the relativistic one (\textsc{kerrbb}). HIDs (absorption corrected) using these fits are presented in Fig. \ref{kerrbb} as filled dots. For the fitting we used a BH mass of 7 $M_{\odot}$ (see \citealt{Orosz2004}; \citealt{Munoz-Darias2008}) and a spin value of $a_*=0.5$ in both cases. Since the \textsc{kerrbb} model includes the inclination as a parameter, we fixed a low inclination value of $40^{\circ}$ for GX~339-4 and $i=70^{\circ}$ was chosen for GRO~J1655-40. 
\item Using the same spectral model (i.e. no further fitting), we computed alternative HIDs by fixing all the spectral parameters but the inclination, which was switched from low to high and vice versa depending on the case. These new simulated HIDs are presented as open circles in Fig.\ref{kerrbb}.
\end{enumerate}
The simulated HID track described by GX 339-4 when the inclination is switched from $40^{\circ}$ to $70^{\circ}$ becomes more similar to that of GRO J1655-40  (Fig. \ref{kerrbb}).  The earlier start of the flux drop and the range of X-ray colours displayed resemble those of the latter source. In the same way, GRO~J1655-40 looks more like GX~339-4 when the viewing angle is changed to $40^{\circ}$, reaching the X-ray colour region ($\leq 0.03$) exclusive to low inclination sources. Although the test is able to reconcile some of the main differences observed, a few discrepancies persist. For instance, the hard-to-soft transitions in GX~339-4 remains more horizontal than that of GRO~J1655-40. Here, we have to bear in mind that the GX~339-4 observations corresponding to the first part of the transition (X-ray colour $\gtrsim$ 0.5) are not sensitive to this test simply because the disc is not detected, and hence fluxes stay the same. However, simulated GRO J1655-40 observations (see open, red circles in the bottom panel of Fig. \ref{kerrbb}) describe an almost horizontal track above X-ray colour $\sim$ 0.1. In light of this, we also observe that the disc flux (and therefore the total), is higher at low inclination. This is expected since the difference in disc projected area would only be compensated by light bending effects for spin parameters close to unity (see above). For this reason, once the disc is clearly detected, it is more dominant when seen at low inclination,  and further dilutes the steep power-law decay that takes places from the middle of the hard-to-soft transition (X-ray colour $\sim 0.2$) onwards. Hence, this effect should contribute to make the transition more horizontal for sources closer to face-on. Note that for the same reason this increase in disc flux should also have a net impact on the colours observed in the soft state. However, this effect (factor two or less) is included in the tolerance range ( $0.002<PLF< 0.05$ in the soft state) allowed to our sample. Indeed, no systematic effect regarding the disc contribution in the softest observations is seen in Fig. \ref{plf}. Finally, we note that limb darkening effects were included in our simulations and found to have negligible impact in our results.\par

We have seen above that the dependence of $T_{obs}$ with the line-of-sight together with simple projection effects are able to explain an important part of the observations. However, other small differences between sources can be also spotted, even if it is not clear whether they are or not related to inclination effects. For instance, from Fig. \ref{hids} it seems that flux spikes are present during the hard-to-soft transitions of at least three high inclination systems, namely, XTE~J1550-560, GRO~J1655-40 and H~1743-322 (also XTE~J1859+226 with $i\gtrsim 60^{\circ}$). We do not have a final answer for this but the earlier detection of the disc at high inclination (due to higher $T_{obs}$) and its larger projected area at low inclination (once it is detected) could play a role in this issue. We also note that \cite{Narayan2005} noticed that high inclination BHT seem to have more complex light curves.\par 
Viewing angle may have also a systematic impact on the power-law emission arising from the inner accretion flow, the geometry of which is unknown. However, we do not find any systematic difference when comparing power-law indexes observed at low and high inclination. On the other hand, we have assumed a constant absorption coefficient ($N_H$) along the outburst. Here, one could speculate about the possibility of variations in the extinction for high inclination sources (e.g. as result of equatorial soft state winds; \citealt{Ponti2012}), that would further attenuate their soft emission. Even though a higher absorption could explain part of what we observe (e.g.  differences in colours), the clear difference in the observed temperature would remain unexplained. As a test, we artificially increased $N_H$ in some observations corresponding to high inclination sources, which did not produce any significant variations in the $T_{obs}$ measured but a worse fit in some of the cases. We also note that \cite{Cabanac2009} studied the evolution of $N_H$ along the outburst for e.g.  GX~339-4 and GRO~J1650-40, finding small variations and similar trends.

\section{Conclusions}
We have shown that inclination has a strong effect in the observed outburst evolution of black hole transients, and consequently alters the hardness-intensity diagrams traditionally used for their study. Quantitatively, high inclination sources are harder (X-ray colour $\geq 0.03$ in our colour scale) and their accretion discs look hotter than those of low inclination systems. The temperature difference can be fully explained by considering relativistic effects predicted decades ago. Assuming that there is no systematic difference in BH mass and accretion rate between low and high inclination systems, our study suggest that at least some of the BH of our sample are spinning. However, other inclination-dependent effects are probably required to explain all the differences observed in the HIDs. Finally, we want to point out that the work presented here might be used in the future to obtain an initial constraint on the orbital inclination based on the shape of the HID track displayed by the system and the range of temperatures and (soft state) colours observed. 
\vspace{2cm}

\noindent  We are grateful to the anonymous referee for useful comments and suggestions that have improved an earlier version of the manuscript. TMD acknowledges funding via an EU Marie Curie Intra-European Fellowship under contract no. 2011-301355. DSP acknowledges financial support provided by the STFC. This project was partially funded by European Research Council Advanced Grant 267697   \textquotedblleft4 pi sky: Extreme Astrophysics with Revolutionary Radio Telescopes\textquotedblright.

\bibliographystyle{mn2e}
\bibliography{/Users/tmd/Dropbox/Libreria.bib} 
\nocite{Dunn2010}

\label{lastpage}
\end{document}